\documentclass[12pt,a4paper]{article}
\usepackage{amsmath}
\usepackage[dvipdfmx]{graphicx}

\begin{document}

\title{
Entropic Approach to Error-Disturbance Tradeoff in Quantum Measurements
}

\author{
Daigo WATANABE and Osamu NARIKIYO\\ 
{\it Department of Physics, Kyushu University, Fukuoka 819-0395, Japan}
}

\maketitle

\begin{abstract}
We discuss the relation between entropic uncertainty relations 
by Buscemi et al. and by Barchielli et al. 

\vskip 15pt 


\vskip 15pt 

\end{abstract}

\vskip 20pt 

The present status of 
the entropic uncertainty relations in quantum measurements 
is almost reviewed by Coles et al.~\cite{1}. 
From the view point of the state-independent error-disturbance tradeoff 
only the work by Buscemi et al.~\cite{2} is described in the section VII-C-1 
in this review. 
More recently a significant development has been made by Barchielli et al.~\cite{3} 
Here we try to put the formulation by Buscemi et al.~\cite{2} 
in the context of Barchielli et al.~\cite{3}

First we reproduce the result by Buscemi et al.~\cite{2} 
after Coles et al.~\cite{1} 
In the sequential measurement 
of the observables ${\hat X}$ and ${\hat Y}$ in the $d$-dimensional Hilbert space 
the error is given by 
\begin{equation}
N(X,M_X) = H(X|M_X) = H(X) - I(X:M_X),
\nonumber 
\end{equation}
and the $\lq\lq$optimized" disturbance by 
\begin{equation}
D(Y,M_Y) = H(Y|M_Y) = H(Y) - I(Y:M_Y),
\nonumber
\end{equation}
where 
$ H(\bullet) $ is the Shannon entropy, 
$ H(\bullet|\bullet) $ is the conditional entropy 
and $ I(\bullet:\bullet) $ is the mutual information. 
Here we have used the static-dynamic isomorphism\cite{1} and these are static expressions. 
The recovery map\cite{1,2} needed for the determination of the disturbance 
is absorbed into the definition of the channel 
complementary to the channel for the determination of the error. 
Thus the effect of the recovery map is not directly seen 
and the resulting bound, $- \log c$, in the following 
is nothing but the basic result obtained by Maassen and Uffink.\cite{4} 
See also the supplementary material of ref. 5 
for the discussion of the $\lq\lq$optimized" disturbance. 
Buscemi et al.~\cite{2} 
assumed the equal weight for the prepared input states so that 
$ H(X) = H(Y) = \log d $. 
Since the mutual information is estimated 
by the information exclusion relation as ref. 1-(235): 
\begin{equation}
I(X:M_X) + I(Y:M_Y) \leq \log(d^2c),
\nonumber
\end{equation}
we obtain 
\begin{equation}
N(X,M_X) + D(Y,M_Y) \geq - \log c, 
\nonumber
\end{equation}
as ref. 1-(375). 
Here $ c = \max_{x,y}| \langle x|y \rangle |^2 $,
with ${\hat X}| x \rangle = x | x \rangle$ and ${\hat Y}| y \rangle = y | y \rangle$. 
The input state of this description is restricted to the uniform ensemble 
of $| x \rangle \langle x |$ or $| y \rangle \langle y |$ 
so that Coles et al.~\cite{1} classified this description 
as a \lq\lq calibration" approach. 
In the following we discuss the case of general input states 
after Barchielli et al.~\cite{3}

A significant progress has been made by Barchielli et al.~\cite{3} 
after the review by Coles et al.~\cite{1}. 
Next we reproduce it in our textbook-level description. 

In the beginning we consider the measurement of the observable $A(x)$ 
where the input quantum state is $\rho$. 
The output quantum state is given by a map $I$ as 
\begin{equation}
{\Large
\mathop{\stackrel{\, \ \rho\ \Longrightarrow\ A \ \Longrightarrow\ I[\rho]}
{\downarrow}}_{\{p(x)\}}
},
\nonumber 
\end{equation}
and the classical output distribution $\{p(x)\}$ is given by 
\begin{equation}
p(x) = {\rm Tr} \{ \rho A(x) \} \equiv A^\rho(x). 
\nonumber
\end{equation}
The map $I$ is composed as 
\begin{equation}
I[\rho] = \sum_x I_x[\rho]. 
\nonumber
\end{equation}
For example, 
$ I_x[\rho] = \sqrt{A(x)} \rho  \sqrt{A(x)} $ 
in the case of L{\"u}ders measurement. 
Similarly the measurement of the observable $B(y)$ is described by a map $J$ as 
\begin{equation}
{\Large
\mathop{\stackrel{\, \ \rho\ \Longrightarrow\ B \ \Longrightarrow\ J[\rho]}
{\downarrow}}_{\{p(y)\}}
},
\nonumber 
\end{equation}
with the classical output 
$ p(y) = {\rm Tr} \{ \rho B(y) \} \equiv B^\rho(y) $. 

Employing the above maps 
we can discuss the straightforward sequential-measurement 
where $B(y)$ is measured after the measurement of $A(x)$. 
The description for $A(x)$ is the same as the above. 
The measurement of $B(y)$ is described as 
\begin{equation}
{\Large
\mathop{\stackrel{\, \ I[\rho]\ \Longrightarrow\ B \ \Longrightarrow\ J[I[\rho]]}
{\downarrow}}_{\{{\tilde p}(y)\}}
},
\nonumber 
\end{equation}
where the quantum input is $I[\rho]$ and the classical output is 
\begin{equation}
{\tilde p}(y) = {\rm Tr} \{ I[\rho] B(y) \}. 
\nonumber
\end{equation}

Introducing the adjoint map $I_x^*$ by 
\begin{equation}
{\rm Tr} \{ I_x[F] G \} = {\rm Tr} \{ F I_x^*[G] \}, 
\nonumber
\end{equation}
for arbitrary observables $F$ and $G$, 
the straightforward sequential-measurement of $A(x)$ and $B(y)$ 
is identified with the measurement of $A(x)$ and $I_x^*[B(y)]$ 
for an equal input $\rho$. 
Using the adjoint map the classical output is rewritten as 
\begin{equation}
{\tilde p}(y) = {\rm Tr} \{ \rho I^*[B(y)] \}, 
\nonumber
\end{equation}
where $ I^*[G] = \sum_x I_x^*[G] $. 

To discuss the general equal-input measurements 
including the straightforward sequential-measurement 
we introduce a bi-observable $M(x,y)$ 
\begin{equation}
{\Large
\mathop{\stackrel{\ \ \rho\ \Longrightarrow\ M \ \Longrightarrow\ K[\rho]}
{\downarrow}}_{\{p(x,y)\}}
},
\nonumber 
\end{equation}
which has two labels $(x,y)$ to classify the outputs. 
This observable describes a measurement 
to obtain an information about both $A(x)$ and $B(y)$. 

If $[A(x),B(y)]=0$ for all $(x,y)$, the construction of $M(x,y)$ is trivial:  
$ M(x,y)=A(x)B(y)=B(y)A(x) $ 
where the marginal of $M(x,y)$ becomes $A(x)$ or $B(y)$ as 
$ \sum _y M(x,y) = A(x) $ or $ \sum_x M(x,y) = B(y) $, 
since 
$ \sum _x A(x) = {\hat 1} $ and $ \sum_y B(y) = {\hat 1} $. 

In the case of the straightforward sequential-measurement described above 
\begin{equation}
K_{xy}[\rho]=J_y[I_x[\rho]]. 
\nonumber
\end{equation}
For arbitrary $\rho$ 
\begin{equation}
{\rm Tr}\{ \rho A(x) \} = {\rm Tr}\{ I_x[\rho] \}, 
\nonumber 
\end{equation}
so that $ A(x) = I_x^*[{\hat 1}] $. 
Similarly 
$ B(y) = J_y^*[{\hat 1}] $ and 
\begin{equation}
M(x,y)=K_{xy}^*[{\hat 1}]=I_x^*[J_y^*[{\hat 1}]]=I_x^*[B(y)]. 
\nonumber
\end{equation}
The marginal of $M(x,y)$ is 
$ \sum_y M(x,y) = I_x^*[{\hat 1}] = A(x) $ or 
$ \sum_x M(x,y) = I^*[B(y)] $. 

If the observable for the second measurement is sharp, 
the equal-input measurement for any bi-observable $M(x,y)$ 
can describe the sequential-measurement as shown in ref. 3-Proposition 5. 
We are interested in the case where the second measurement is sharp. 

Since an exact measurement for both $A(x)$ and $B(y)$ is impossible in general, 
real measurements should be approximations. 
Although we are interested in 
the error-disturbance tradeoff in a sequential measurement, 
it can be identified with that in an equal-input measurement as discussed above. 
Then we consider an equal-input measurement 
described by the bi-observable $M(x,y)$ 
which gives an approximation
for the distributions $A^\rho \equiv \{ A^\rho(x) \}$ 
and $B^\rho \equiv \{ B^\rho(y) \}$. 
The approximated distribution compared with $A^\rho$ is the set of 
$ M_1^\rho(x) = {\rm Tr}\{ \rho M_1(x) \} $ 
where 
$ M_1(x) = \sum_y M(x,y) $. 
Similarly the approximated distribution compared with $B^\rho$ is the set of 
$ M_2^\rho(y) = {\rm Tr}\{ \rho M_2(y) \} $ 
where 
$ M_2(y) = \sum_x M(x,y) $. 

An explicit form of $M(x,y)$ for the case of qubit measurements is as follows. 
The construction of the bi-observable for qubit experiments 
by Barchielli et al.~\cite{3} is done on the basis of a symmetry argument. 
However, experimental implementations become transparent 
by the description of approximation schemes such as 
the stochastic mixing~\cite{5} and the misalignment.~\cite{6}
Anyway, 
the bi-observable for mutually orthogonal observables is explicitly obtained as 
\begin{equation}
M(x,y)={1 \over 4}
       [ {\hat 1} + {x \over \sqrt{2}}{\hat X} 
                  + {y \over \sqrt{2}}{\hat Y} ].
\nonumber 
\end{equation}
For this bi-observable the mini-max estimation of the error plus disturbance 
can be carried out analytically.~\cite{3} 
By the discussion in ref. 3, however, 
neither of these approximation schemes~\cite{5,6} seems to work 
for the non-orthogonal case. 

The difference between the distributions $p$ and $q$ 
can be measured by the relative entropy
\begin{equation}
S(p||q)=\sum_x p(x) [ \log p(x) - \log q(x) ].
\nonumber
\end{equation}
Thus the difference between ideal and real measurements can be measured by 
$S(A^\rho||M_1^\rho)$ and $S(B^\rho||M_2^\rho)$. 

A tradeoff is realized by optimizing the sum 
$S(A^\rho||M_1^\rho) + S(B^\rho||M_2^\rho)$. 
Here we do not define the error and the disturbance individually 
but consider the error plus the disturbance. 

The optimization is done by the mini-max principle~\cite{7} 
and the estimate for the error plus disturbance is given by 
\begin{equation}
\min_{M} [ \max_{\rho} \{ S(A^\rho || M^\rho_1) + S(B^\rho || M^\rho_2)  \} ].
\nonumber
\end{equation}
Consequently our estimation for the error plus disturbance is an operational one 
different from the explicit bound as $- \log c$. 

Now we have finished our textbook-level reproduction 
of the result by Barchielli et al.~\cite{3} 
and are ready to put the formulation by Buscemi et al.~\cite{2} 
in the context of Barchielli et al.~\cite{3} 
First we generalize the way~\cite{8} of calculating the conditional entropy 
for a single-observable $M_m$ to describe general measurement situations. 
Then we apply it to the bi-observable. 

The conditional entropy is calculated as 
\begin{equation}
H(A|M) = - \sum_m p(m) \sum_a p(a|m) \log p(a|m). 
\nonumber 
\end{equation}
The input quantum state is 
\begin{equation}
\rho = \sum_a p(a) | a \rangle \langle a |,  
\nonumber 
\end{equation}
and the output classical distribution is 
\begin{equation}
p(m) = {\rm Tr} \{ M_m \rho \} 
     = \sum_a p(a) {\rm Tr} \{ M_m | a \rangle \langle a | \}. 
\nonumber 
\end{equation}
On the other hand, $p(m)$ is the marginal of $p(m,a)$; 
$ p(m) = \sum_a p(m,a) = \sum_a p(a) p(m|a) $  
where 
the joint probability $p(m,a)$ and 
the conditional probability $p(m|a)$ is related by $p(m,a)=p(m|a)p(a)$. 
The above two expressions for $p(m)$ show 
\begin{equation}
p(m|a) = {\rm Tr} \{ M_m | a \rangle \langle a | \}. 
\nonumber 
\end{equation}
Thus we obtain the master relation 
\begin{equation}
p(m,a) = p(a) {\rm Tr} \{ M_m | a \rangle \langle a | \}. 
\nonumber 
\end{equation}
From this master relation 
we can derive $p(a|m)$ as 
$ p(a|m) = p(m,a) / p(m) $, 
since $p(m,a)=p(a|m)p(m)$. 
Here  
$ p(m) = \sum_a p(m,a) $. 

The conditional entropy is also written as 
\begin{equation}
H(A|M) = - \sum_m \sum_a p(m,a) \log p(a|m). 
\nonumber 
\end{equation}

Buscemi et al.~\cite{2} 
measure the error and the disturbance by the conditional entropy. 
Their input was restricted to the uniform ensemble as already discussed. 
Here we employ the scheme by Barchielli et al.~\cite{3} 
for general inputs. 
Consequently we can put the formulation by Buscemi et al.~\cite{2} 
in the context of Barchielli et al.~\cite{3} 
where 
the error plus disturbance is characterized by
\begin{equation}
\min_{M} [ \max_{\rho} \{ H(A^\rho | M^\rho_1) + H(B^\rho | M^\rho_2)  \} ]. 
\nonumber
\end{equation}
Here the conditional entropy is calculated by 
\begin{equation}
H(A^\rho | M^\rho_1) = - \sum_x \sum_{x'} p(x,x') \log p(x|x'). 
\nonumber
\end{equation}
The master, joint probability, is given as 
\begin{equation}
p(x,x') = A^\rho(x) {\rm Tr} \{ M_1(x') | x \rangle \langle x | \}. 
\nonumber
\end{equation}
From the master 
the marginal is derived as 
$ p(x') = \sum_x p(x,x') $  
and the conditional probability is derived as  
$ p(x|x') = p(x,x') / p(x') $.

\end{document}